\documentclass[prd,showpacs,floatfix,twocolumn,,amsmath,amssymb,floatfix]{revtex4}
\usepackage{graphicx,color,dcolumn,booktabs,bm}
\usepackage{longtable,lscape}
\usepackage{txfonts}
\usepackage{overpic}
\usepackage{amssymb}
\usepackage{indentfirst}
\usepackage{feynmf}   
\usepackage{slashed}  
\usepackage{cases}
\usepackage{color}
\usepackage{multirow}
\usepackage{graphicx,color,dcolumn,booktabs,bm}

\graphicspath{{Figures/}} %

\graphicspath{{Figures/}} %

\usepackage[colorlinks, citecolor=blue,anchorcolor=red,menucolor=red, linkcolor=red,filecolor=red,runcolor=red,urlcolor=blue,frenchlinks=red]{hyperref}

\allowdisplaybreaks

\begin{document}


\title{Can the newly $P_{cs}(4459)$ be a strange hidden-charm $\Xi_c\bar{D}^{*}$ molecular pentaquarks?}
\author{Rui Chen$^{1,2}$}
\email{chen$_$rui@pku.edu.cn}
\affiliation{
$^1$Center of High Energy Physics, Peking University, Beijing
100871, China\\
$^2$ School of Physics and State Key Laboratory of Nuclear Physics and Technology, Peking University, Beijing 100871, China}

\date{\today}

\begin{abstract}
  Stimulated by the $P_{cs}(4459)$ reported by the LHCb Collaboration, we perform a single $\Xi_c\bar{D}^*$ channel and a coupled $\Xi_c\bar{D}^*/\Xi_c^*\bar{D}/\Xi_c^{\prime}\bar{D}^*/\Xi_c^*\bar{D}^*$ channel analysis by using a one-boson-exchange model. Our results indicate that the newly $P_{cs}(4459)$ cannot be a pure $\Xi_c\bar{D}^*$ molecular state, but a coupled $\Xi_c\bar{D}^*/\Xi_c^*\bar{D}/\Xi_c^{\prime}\bar{D}^*/\Xi_c^*\bar{D}^*$ bound state with $I(J^P)=0(3/2^-)$, where the $\Xi_c\bar{D}^*$ and $\Xi_c^*\bar{D}$ components are dominant. Meanwhile, we find the interactions from the $\Xi_c^{\prime}\bar{D}^*$ system with $0(1/2^-)$, the $\Xi_c^{*}\bar{D}$ system with $1(3/2^-)$, and the $\Xi_c^{*}\bar{D}^*$ system with $1(1/2^-)$ are strongly attractive, where one can expect possible strange hidden-charm molecular or resonant structures near the these thresholds with the assigned quantum numbers.

\end{abstract}

\maketitle

\section{introduction}\label{sec1}

In 2019, the LHCb Collaboration updated the observations in the $\Lambda_b^0\to J/\psi pK^-$ process by using more data \cite{Aaij:2019vzc}, they not only discovered a new narrow pentaquark state, $P_c(4312)^+$, but also found that the $P_c(4450)$ reported previously \cite{Aaij:2015tga} consists of two narrow overlapping peaks, $P_c(4440)^+$ and $P_c(4457)^+$.

The discovery of $P_c$ states has sparked an enormous interest in the study of multiquark systems and exotic hadrons. Several possible explanations have been put forward in the literatures: the molecular states \cite{Chen:2015loa,Chen:2015moa,Roca:2015dva,Chen:2019asm,Liu:2019tjn,He:2019ify,Meng:2019ilv,Burns:2019iih}, the compact pentaquark states \cite{Maiani:2015vwa,Lebed:2015tna,Weng:2019ynv,Ali:2019npk,Giron:2019bcs,Cheng:2019obk}, and the kinematical effects \cite{Guo:2015umn,Liu:2015fea} (see review papers \cite{Chen:2016qju,Liu:2019zoy,Brambilla:2019esw,Guo:2017jvc,Esposito:2016noz,Hosaka:2016pey} for more details). In fact, before the observations of the $P_c$ states, the hidden-charm pentaquarks were predicted in Refs. \cite{Yang:2011wz,Wu:2010jy,Wang:2011rga,Karliner:2015ina,Li:2014gra}.

Among the different interpretations to these $P_c$ states, the hadronic molecular state assignments to them are the most popular one. The masses of the $P_c$ states are close to the thresholds of a charmed baryon and an anticharmed meson, which is the main reason why the hadronic molecular state assignments to them were proposed. For example, $P_c(4312)$, $P_c(4440)$, and $P_c(4457)$ are regards as the hidden-charm molecular pentauqarks, they are mainly composed by the $\Sigma_c\bar{D}$ state with $I(J^P)=1/2(1/2^-)$, the $\Sigma_c\bar{D}^*$ state with $1/2(1/2^-)$ and $1/2(3/2^-)$, respectively \cite{Chen:2019asm}.

Very recently, the LHCb Collaboration reported an evidence of the $P_{cs}(4459)$ in $\Xi_b^-\to J/\psi\Lambda K^-$ with a 3.1 $\sigma$ statistical significant \cite{wang:talk}. Its mass and width are
\begin{eqnarray*}
M=4458.8\pm2.9_{-1.2}^{+4.7} \text{MeV}, \quad \Gamma= 17.3\pm6.5_{-5.7}^{+8.0} \text{MeV},
\end{eqnarray*}
respectively. Its spin-parity was not determined yet. According to the decay final states $J/\psi\Lambda$, the $P_{cs}(4459)$ is a strange hidden-charm pentaquark containing $c\bar{c}sud$ valence quark components. Since its mass is just below the $\Xi_c\bar{D}^*$ threshold with around 19 MeV, whether the newly $P_{cs}(4459)$ can be explained as a strange hidden-charm $\Xi_c\bar{D}^*$ molecule is open to discuss \cite{Chen:2020uif,Peng:2020hql,Wang:2020eep}, i.e., Chen {\it et al.}'s results supported the $P_{cs}(4459)$ as the $\Xi_c\bar{D}^*$ molecule of either $J^P=1/2^-$ or $3/2^-$ after adopted the QCD sum rules \cite{Chen:2020uif}.

In fact, many groups have predicted the existence of the strange hidden-charm pentaquarks \cite{Wang:2019nvm,Chen:2016ryt,Anisovich:2015zqa,Wang:2015wsa,Feijoo:2015kts,Lu:2016roh,Xiao:2019gjd,Zhang:2020cdi,Shen:2020gpw}, and proposed to search for the $P_{cs}$ states in the $\Lambda_b(\Xi_b)\to J/\psi\Lambda K(\eta)$ \cite{Lu:2016roh,Feijoo:2015kts,Chen:2015sxa}. Especially, Wang {\it et al.} predicted two isoscalar $\Xi_c\bar{D}^{*}$ molecular states with the chiral effective field theory \cite{Wang:2019nvm}, the masses of $\Xi_c\bar{D}^{*}$ molecules with $J^P=1/2^-$ and $3/2^-$ are $4456.9_{-3.3}^{+3.2}$ MeV and $4463.0_{-3.0}^{+2.8}$ MeV, respectively.

As we mentioned, the long range interaction from one pseudoscalar meson exchange is suppressed in the single $\Xi_c\bar{D}^*$ system \cite{Chen:2016ryt}. In this work, we will give a more comprehensive and systematic investigation of the molecular explanation of the $P_{cs}(4459)$ after absorbing more effects. And we still adopt the one-boson-exchange (OBE) model (including the $\pi$, $\sigma$, $\eta$, $\rho$, and $\omega$ exchanges), and adopt the $S-D$ wave mixing and the coupled-channel effect. By studying the hadronic molecular state assignment to the $P_{cs}(4459)$, we want to further identify this molecular pentaquark configuration, especially the corresponding spin-parity quantum numbers.

This paper is organized as follows. After this introduction, we illustrate the deducing of the OBE effective potentials in Sec.~\ref{sec2}. The corresponding numerical results for the single $\Xi_c\bar{D}^*$ channel case and the coupled $\Xi_c\bar{D}^*/\Xi_c^*\bar{D}/\Xi_c'\bar{D}^*/\Xi_c^*\bar{D}^*$ case are given in Sec.~\ref{sec3} and Sec. \ref{sec4}, respectively. The paper ends with a summary in Sec. \ref{sec4}.

\section{one-boson-exchange effective potentials}\label{sec2}

According to the heavy quark symmetry and chiral symmetry \cite{Yan:1992gz,Wise:1992hn,Burdman:1992gh,Casalbuoni:1996pg,Falk:1992cx,Liu:2011xc}, the relevant effective Lagrangians are constructed as
\begin{eqnarray}
\mathcal{L}_{H} &=& g_S\langle \bar{H}_a^{(\bar{Q})}\sigma H_b^{(\bar{Q})}\rangle
  +ig\langle \bar{H}_a^{(\bar{Q})}\gamma_{\mu}A_{ab}^{\mu}\gamma_5H_b^{(\bar{Q})}\rangle\nonumber\\
  &&-i\beta\langle \bar{H}_a^{(\bar{Q})}v_{\mu}\left(\mathcal{V}_{ab}^{\mu}-\rho_{ab}^{\mu}\right)
  H_b^{(\bar{Q})}\rangle\nonumber\\
  &&+i\lambda\langle \bar{H}_a^{(\bar{Q})}\sigma_{\mu\nu}F^{\mu\nu}(\rho)H_b^{(\bar{Q})}\rangle,\label{lag1}\\
\mathcal{L}_{\mathcal{B}_{\bar{3}}} &=& l_B\langle\bar{\mathcal{B}}_{\bar{3}}\sigma\mathcal{B}_{\bar{3}}\rangle
          +i\beta_B\langle\bar{\mathcal{B}}_{\bar{3}}v^{\mu}(\mathcal{V}_{\mu}-\rho_{\mu})\mathcal{B}_{\bar{3}}\rangle,\label{lag2}\\
\mathcal{L}_{\mathcal{B}_{6}} &=&  l_S\langle\bar{\mathcal{S}}_{\mu}\sigma\mathcal{S}^{\mu}\rangle
         -\frac{3}{2}g_1\varepsilon^{\mu\nu\lambda\kappa}v_{\kappa}\langle\bar{\mathcal{S}}_{\mu}A_{\nu}\mathcal{S}_{\lambda}\rangle\nonumber\\
         &&+i\beta_{S}\langle\bar{\mathcal{S}}_{\mu}v_{\alpha}\left(\mathcal{V}_{ab}^{\alpha}-\rho_{ab}^{\alpha}\right) \mathcal{S}^{\mu}\rangle
         +\lambda_S\langle\bar{\mathcal{S}}_{\mu}F^{\mu\nu}(\rho)\mathcal{S}_{\nu}\rangle,\nonumber\\\\
\mathcal{L}_{\mathcal{B}_{\bar{3}}\mathcal{B}_6} &=& ig_4\langle\bar{\mathcal{S}^{\mu}}A_{\mu}\mathcal{B}_{\bar{3}}\rangle
         +i\lambda_I\varepsilon^{\mu\nu\lambda\kappa}v_{\mu}\langle \bar{\mathcal{S}}_{\nu}F_{\lambda\kappa}\mathcal{B}_{\bar{3}}\rangle+h.c..\label{lag3}
\end{eqnarray}
Here, the multiplet field $H^{(\bar{Q})}$ is composed by the pseudoscalar meson $\tilde{\mathcal{P}}=\left(\bar{D}^0,\,D^-\right)^T$ and the vector meson $\tilde{\mathcal{P}}^*=\left(\bar{D}^{*0},\,D^{*-}\right)^T$, while its conjugate field of $\bar{H}^{(\bar{Q})}$ satisfies $\bar{H}^{(\bar{Q})}=\gamma_0H^{(\bar{Q})\dag}\gamma_0$. And $\mathcal{S}$ is defined as a superfield, which includes $\mathcal{B}_6$ with $J^P=1/2^+$ and $\mathcal{B}^*_6$ with $J^P=3/2^+$ in the $6_F$ flavor representation. Their expressions read as
\begin{eqnarray}
H^{(\bar{Q})} &=& [\tilde{\mathcal{P}}^{*\mu}\gamma_{\mu}-\tilde{\mathcal{P}}\gamma_5]\frac{1-\rlap\slash v}{2},\\
\mathcal{S}_{\mu} &=&
-\sqrt{\frac{1}{3}}(\gamma_{\mu}+v_{\mu})\gamma^5\mathcal{B}_6
       +\mathcal{B}_{6\mu}^*.
\end{eqnarray}
The expressions of the axial current and the vector current denote as
\begin{eqnarray*}
A_{\mu} &=& \frac{1}{2}(\xi^{\dag}\partial_{\mu}\xi-\xi\partial_{\mu}\xi^{\dag})=\frac{i}{f_{\pi}}
\partial_{\mu}\mathbb{P}+\ldots,\\
\mathcal{V}_{\mu} &=&
\frac{1}{2}(\xi^{\dag}\partial_{\mu}\xi-\xi\partial_{\mu}\xi^{\dag})
 =\frac{i}{2f_{\pi}^2}\left[\mathbb{P},\partial_{\mu}\mathbb{P}\right]+\ldots,
\end{eqnarray*}
respectively, with $\xi=\text{exp}(i\mathbb{P}/f_{\pi})$ and the pion decay constant $f_{\pi}=132$ MeV. $\rho_{ba}^{\mu}=ig_V\mathbb{V}_{ba}^{\mu}/\sqrt{2}$, $F^{\mu\nu}(\rho)=\partial^{\mu}\rho^{\nu}-\partial^{\nu}\rho^{\mu}
+\left[\rho^{\mu},\rho^{\nu}\right]$. $\mathbb{P}$ and $\mathbb{V}$ stand for the isoscalar and vector matrices, respectively. In the above formulas, the matrices $\mathcal{B}_{\bar{3}}$, $\mathcal{B}_6^{(*)}$, $\mathbb{P}$, and $\mathcal{V}$ are written as
\begin{eqnarray*}
\mathcal{B}_{\bar{3}} &=& {\left(\begin{array}{ccc}
         0    &\Lambda_c^+       &\Xi_c^+\\
        -\Lambda_c^+      &0     &\Xi_c^0\\
       -\Xi_c^+      &-\Xi_c^0    &0
                \end{array}\right),}\\
\mathcal{B}_6^{(*)} &=& {\left(\begin{array}{ccc}
         \Sigma_c^{(*)++}              &\frac{1}{\sqrt{2}}\Sigma_c^{(*)+}    &\frac{1}{\sqrt{2}}\Xi_c^{(',*)+}\\
         \frac{1}{\sqrt{2}}\Sigma_c^{(*)+}      &\Sigma_c^{(*)0}       &\frac{1}{\sqrt{2}}\Xi_c^{(',*)0}\\
          \frac{1}{\sqrt{2}}\Xi_c^{(',*)+}     &\frac{1}{\sqrt{2}}\Xi_c^{(',*)0}      &\Omega_c^{(*)0}\\
                \end{array}\right)},\\
\mathbb{P} &=& {\left(\begin{array}{ccc}
       \frac{\pi^0}{\sqrt{2}}+\frac{\eta}{\sqrt{6}} &\pi^+      &K^+\\
       \pi^-       &-\frac{\pi^0}{\sqrt{2}}+\frac{\eta}{\sqrt{6}}     &K^0\\
       K^-         &\bar{K}^0      &-\frac{2}{\sqrt{6}}\eta
               \end{array}\right)},\\
{V} &=& {\left(\begin{array}{ccc}
\frac{\rho^0}{\sqrt{2}}+\frac{\omega}{\sqrt{2}}  &\rho^+      &K^{*+}\\
\rho^- &-\frac{\rho^0}{\sqrt{2}}+\frac{\omega}{\sqrt{2}}      &K^{*0}\\
K^{*-}      &\bar{K}^{*0}    &\phi
\end{array}\right)}.
\end{eqnarray*}

By expanding Eqs. (\ref{lag1})-(\ref{lag3}), one can further get
\begin{eqnarray}
\mathcal{L}_{\tilde{\mathcal{P}}^{(*)}\tilde{\mathcal{P}}^{(*)}\sigma} &=& -2g_S\tilde{\mathcal{P}}_b^{\dag}\tilde{\mathcal{P}}_b\sigma-2g_S\tilde{\mathcal{P}}_b^{*}\cdot\tilde{\mathcal{P}}_b^{*\dag}\sigma,\\
\mathcal{L}_{\tilde{\mathcal{P}}^{(*)}\tilde{\mathcal{P}}^{(*)}\mathbb{P}} &=& \frac{2g}{f_{\pi}}\left(\tilde{\mathcal{P}}_{a\lambda}^{*\dag}\tilde{\mathcal{P}}
+\tilde{\mathcal{P}}_{a}^{\dag}\tilde{\mathcal{P}}^*_{b\lambda}\right)
\partial^{\lambda}\mathbb{P}_{ab}\nonumber\\
&&+i\frac{2g}{f_{\pi}}v^{\alpha}\varepsilon_{\alpha\mu\nu\lambda}
\tilde{\mathcal{P}}_{a}^{*\mu\dag}
\tilde{\mathcal{P}}_b^{*\lambda}\partial^{\nu}\mathbb{P}_{ab},\\
\mathcal{L}_{\tilde{\mathcal{P}}^{(*)}\tilde{\mathcal{P}}^{(*)}\mathbb{V}} &=& \sqrt{2}\beta g_V\tilde{\mathcal{P}}_a^{\dag}\tilde{\mathcal{P}}_b v\cdot\mathbb{V}_{ab}\nonumber\\
&&-2\sqrt{2}\lambda g_Vv^{\lambda}\varepsilon_{\lambda\mu\alpha\beta}\left(\tilde{\mathcal{P}}^{*\mu\dag}_a
\tilde{\mathcal{P}}_b+\tilde{\mathcal{P}}_a^{\dag}\tilde{\mathcal{P}}_b^{*\mu}\right)
    \partial^{\alpha}\mathbb{V}^{\beta}_{ab}\nonumber\\
    &&-\sqrt{2}\beta g_V\tilde{\mathcal{P}}_a^{*\dag}\cdot\tilde{\mathcal{P}}_b^*v\cdot\mathbb{V}_{ab}\nonumber\\
   &&-i2\sqrt{2}\lambda g_V\tilde{\mathcal{P}}_a^{*\mu\dag}\tilde{\mathcal{P}}_b^{*\nu}
   \left(\partial_{\mu}\mathbb{V}_{\nu}-\partial_{\nu}\mathbb{V}_{\mu}\right),\\
\mathcal{L}_{\mathcal{B}_{\bar{3}}\mathcal{B}_{\bar{3}}\sigma} &=& l_B\langle \bar{\mathcal{B}}_{\bar{3}}\sigma\mathcal{B}_{\bar{3}}\rangle,\\
\mathcal{L}_{\mathcal{B}_{6}^{(*)}\mathcal{B}_{6}^{(*)}\sigma} &=& -l_S\langle\bar{\mathcal{B}}_6\sigma\mathcal{B}_6\rangle+l_S\langle\bar{\mathcal{B}}_{6\mu}^{*}\sigma\mathcal{B}_6^{*\mu}\rangle\nonumber\\
       &&-\frac{l_S}{\sqrt{3}}\langle\bar{\mathcal{B}}_{6\mu}^{*}\sigma\left(\gamma^{\mu}
       +v^{\mu}\right)\gamma^5\mathcal{B}_6\rangle+h.c.,\\
\mathcal{L}_{\mathcal{B}_{\bar{3}}\mathcal{B}_{\bar{3}}{V}} &=& \frac{1}{\sqrt{2}}\beta_Bg_V\langle\bar{\mathcal{B}}_{\bar{3}}v\cdot{V}\mathcal{B}_{\bar{3}}\rangle,\\
\mathcal{L}_{\mathcal{B}_6^{(*)}\mathcal{B}_6^{(*)}{P}} &=&
        i\frac{g_1}{2f_{\pi}}\varepsilon^{\mu\nu\lambda\kappa}v_{\kappa}\langle\bar{\mathcal{B}}_6
        \gamma_{\mu}\gamma_{\lambda}\partial_{\nu}{P}\mathcal{B}_6\rangle\nonumber\\
      &&+i\frac{\sqrt{3}}{2}\frac{g_1}{f_{\pi}}v_{\kappa}\varepsilon^{\mu\nu\lambda\kappa}
      \langle\bar{\mathcal{B}}_{6\mu}^*\partial_{\nu}{P}{\gamma_{\lambda}\gamma^5}
      \mathcal{B}_6\rangle+h.c.\nonumber\\
      &&-i\frac{3g_1}{2f_{\pi}}\varepsilon^{\mu\nu\lambda\kappa}v_{\kappa}\langle
\bar{\mathcal{B}}_{6\mu}^{*}\partial_{\nu} {P}\mathcal{B}_{6\lambda}^*\rangle,\\
\mathcal{L}_{\mathcal{B}_6^{(*)}\mathcal{B}_6^{(*)} {V}} &=& -\frac{\beta_Sg_V}{\sqrt{2}}\langle\bar{\mathcal{B}}_6v\cdot{V}\mathcal{B}_6\rangle\nonumber\\
    &&-i\frac{\lambda g_V}{3\sqrt{2}}\langle\bar{\mathcal{B}}_6\gamma_{\mu}\gamma_{\nu}
    \left(\partial^{\mu} {V}^{\nu}-\partial^{\nu} {V}^{\mu}\right)
    \mathcal{B}_6\rangle\nonumber\\
   &&-\frac{\beta_Sg_V}{\sqrt{6}}\langle\bar{\mathcal{B}}_{6\mu}^*v\cdot {V}\left(\gamma^{\mu}+v^{\mu}\right)\gamma^5\mathcal{B}_6\rangle\nonumber\\
    &&-i\frac{\lambda_Sg_V}{\sqrt{6}}\langle\bar{\mathcal{B}}_{6\mu}^*
    \left(\partial^{\mu} {V}^{\nu}-\partial^{\nu} {V}^{\mu}\right)
    \left(\gamma_{\nu}+v_{\nu}\right)\gamma^5\mathcal{B}_6\rangle\nonumber\\
    &&+\frac{\beta_Sg_V}{\sqrt{2}}\langle\bar{\mathcal{B}}_{6\mu}^*v\cdot {V}\mathcal{B}_6^{*\mu}\rangle\nonumber\\
    &&+i\frac{\lambda_Sg_V}{\sqrt{2}}\langle\bar{\mathcal{B}}_{6\mu}^*
    \left(\partial^{\mu} {V}^{\nu}-\partial^{\nu} {V}^{\mu}\right)
    \mathcal{B}_{6\nu}^*\rangle+h.c.,\\
\mathcal{L}_{\mathcal{B}_{\bar{3}}\mathcal{B}_6^{(*)}{V}} &=&
       -\frac{\lambda_Ig_V}{\sqrt{6}}\varepsilon^{\mu\nu\lambda\kappa}v_{\mu}\langle \bar{\mathcal{B}}_6\gamma^5\gamma_{\nu}
        \left(\partial_{\lambda} {V}_{\kappa}-\partial_{\kappa} {V}_{\lambda}\right)\mathcal{B}_{\bar{3}}\rangle\nonumber\\
          &&-\frac{\lambda_Ig_V}{\sqrt{2}}\varepsilon^{\mu\nu\lambda\kappa}v_{\mu}\langle \bar{\mathcal{B}}_{6\nu}^*
          \left(\partial_{\lambda}{V}_{\kappa}-\partial_{\kappa}{V}_{\lambda}\right)\mathcal{B}_{\bar{3}}\rangle+h.c.,\nonumber\\\\
\mathcal{L}_{\mathcal{B}_{\bar{3}}\mathcal{B}_6^{(*)} {P}} &=& -\sqrt{\frac{1}{3}}\frac{g_4}{f_{\pi}}\langle\bar{\mathcal{B}}_6\gamma^5\left(\gamma^{\mu}
   +v^{\mu}\right)\partial_{\mu}{P}\mathcal{B}_{\bar{3}}\rangle\nonumber\\
           &&-\frac{g_4}{f_{\pi}}\langle\bar{\mathcal{B}}_{6\mu}^*\partial^{\mu} {P}\mathcal{B}_{\bar{3}}\rangle+h.c.,
\end{eqnarray}
which will be applied to the deduction of scattering amplitudes.

For the $\pi$ exchange couplings, they are extracted from the decay width of $D^*\to D\pi$, $\Sigma_c\to\Lambda_c\pi$, and $\Sigma_c^*\to\Lambda_c\pi$ \cite{Isola:2003fh,Liu:2011xc,pdg}. The remaining coupling constants relevant to the heavy baryons can be estimated by borrowing the nucleon-nucleon interaction in the quark level \cite{Liu:2011xc}. $g_s=\tilde g/2\sqrt{6}$, $\tilde g$ is the coupling for the process $D(0^+)\to D(0^-)+\pi$ \cite{Bardeen:2003kt}. $\beta$ is fixed as $\beta=$0.9 according to vector meson dominance \cite{Isola:2003fh}. And $\lambda$ is determined through a comparison of the form factor between the theoretical calculation from the light cone sum rule and lattice QCD \cite{Isola:2003fh}. Their values are collected in Table \ref{coupling}.
\renewcommand\tabcolsep{0.12cm}
\renewcommand{\arraystretch}{1.5}
\begin{table}[!htbp]
\caption{Coupling constants adopted in our calculation.\label{coupling}}
\begin{tabular}{ccccccccccc}
\toprule[1pt]
    $l_B$     &$\beta_Bg_V$     &$l_S$    &$g_1$       &$\lambda_Sg_V~(\text{GeV}^{-1})$     &$\beta g_V$\\
    $-$3.65    &$-$6.00             &7.30      &1.00    &19.20      &5.22    \\\hline
    $\beta_Sg_V$      &$g_S$       &$g_4$      &$\lambda_Ig_V~(\text{GeV}^{-1})$     &$\lambda g_V~(\text{GeV}^{-1})$\\
    12.00     &0.76      &1.06       &$-$6.80                     &3.25          \\\bottomrule[1pt]
\end{tabular}
\end{table}

In a Breit approximation, the effective potentials in the momentum space can be related to the corresponding scattering amplitudes, i.e.,
\begin{eqnarray}\label{breit}
\mathcal{V}_{E}^{h_1h_2\to h_3h_4}(\bm{q}) &=&
          -\frac{\mathcal{M}(h_1h_2\to h_3h_4)}
          {\sqrt{\prod_i2M_i\prod_f2M_f}},
\end{eqnarray}
where $M_i$ and $M_f$ stand for the masses of the initial states ($h_1$, $h_2$) and final states ($h_3$, $h_4$), respectively. After performing a Fourier transformation, we obtain the effective potential in the coordinate space $\mathcal{V}(\bm{r})$,
\begin{eqnarray}
\mathcal{V}_{E}^{h_1h_2\to h_3h_4}(\bm{r}) =
          \int\frac{d^3\bm{q}}{(2\pi)^3}e^{i\bm{q}\cdot\bm{r}}
          \mathcal{V}_{E}^{h_1h_2\to h_3h_4}(\bm{q})\mathcal{F}^2(q^2,m_E^2).\nonumber
\end{eqnarray}
Here, we introduce a monopole form factor $\mathcal{F}(q^2,m_E^2)=(\Lambda^2-m_E^2)/(\Lambda^2-q^2)$ at every interactive vertex, it expresses the off-shell effect of the exchanged boson. $\Lambda$, $m_E$, and $q$ are the cutoff, mass, and four-momentum of the exchanged meson, respectively. As a phenomenological parameter, the reasonable value of cutoff is taken around 1 GeV \cite{Tornqvist:1993ng,Tornqvist:1993vu}.

For the $S-$wave $\Xi_c\bar{D}^*$ system, its spin-parity can be either $J^P=1/2^-$ or $3/2^-$. When we consider the $S-D$ wave mixing effect, the spin-orbit wave functions are
\begin{eqnarray}\label{spinwave}
\begin{array}{ccccc}
J^P=\frac{1}{2}^-:    &|{}^2\mathbb{S}_{\frac{1}{2}}\rangle, &|{}^4\mathbb{D}_{\frac{1}{2}}\rangle &|{}^6\mathbb{D}_{\frac{1}{2}}\rangle,\\
J^P=\frac{3}{2}^-:    &|{}^4\mathbb{S}_{\frac{3}{2}}\rangle,   &|{}^2\mathbb{D}_{\frac{3}{2}}\rangle,    &|{}^4\mathbb{D}_{\frac{3}{2}}\rangle  &|{}^6\mathbb{D}_{\frac{3}{2}}\rangle.
\end{array}
\end{eqnarray}
The flavor wave functions $|I,I_3\rangle$ of these discussed systems are
\begin{eqnarray}\left.\begin{array}{cl}
\Xi_c^{(\prime,*)}\bar{D}^*:       &\left\{\begin{array}{l}|1,1\rangle=|\Xi_c^{(\prime,*)+}\bar{D}^{*0}\rangle,\\
                                              |1,0\rangle=\frac{1}{\sqrt{2}}\left(|\Xi_c^{(\prime,*)+}D^{*-}\rangle+|\Xi_c^{(\prime,*)0}\bar{D}^{*0}\rangle\right),\\
                                              |1,-1\rangle=|\Xi_c^{(\prime,*)0}D^{*-}\rangle,
                       \end{array}\right.\\
                      &|0,0\rangle=\frac{1}{\sqrt{2}}\left(|\Xi_c^{(\prime,*)+}D^{*-}\rangle-|\Xi_c^{(\prime,*)0}\bar{D}^{*0}\rangle\right),
\end{array}\right.
\end{eqnarray}
where $I$ and $I_3$ are the isospin and its third component of the systems, respectively,

\section{A single $\Xi_c\bar{D}^*$ analysis}\label{sec3}

With the above preparations, the OBE effective potential for the single $\Xi_c\bar{D}^*$ system is written as
\begin{eqnarray}
{V}_{\Xi_c\bar{D}^*} &=& 2l_Bg_s\bm{\epsilon}_2\cdot\bm{\epsilon_4^{\dag}}Y(\Lambda,m_{\sigma},r)\nonumber\\
&&-\frac{\mathcal{G}(I)}{4}\beta\beta_Bg_v^2
\bm{\epsilon}_2\cdot\bm{\epsilon_4^{\dag}}Y(\Lambda,m_{\rho},r)\nonumber\\
&&-\frac{1}{4}\beta\beta_Bg_v^2
\bm{\epsilon}_2\cdot\bm{\epsilon_4^{\dag}}Y(\Lambda,m_{\omega},r),\label{pot1}
\end{eqnarray}
where $\mathcal{G}(I)$ is the isospin factor, its value is taken as $1$ for the isospin-$1$ system, and $-3$ for the isospin-$0$ system. The function $Y(\Lambda,m,{r})$ denotes
\begin{eqnarray}
Y(\Lambda,m,{r}) &=&\frac{1}{4\pi r}(e^{-mr}-e^{-\Lambda r})-\frac{\Lambda^2-m^2}{8\pi \Lambda}e^{-\Lambda r}.
\end{eqnarray}

When producing the numerical calculations, the spin-spin operator $\mathcal{O}^{i,j}$ should be replaced by a serial of matrix elements $\langle{}^{2s'+1}L'_{J'}|\mathcal{O}^{i,j}|{}^{2s+1}L_{J}\rangle$, the $|{}^{2s+1}L_{J}\rangle$ and $|{}^{2s'+1}L'_{J'}\rangle$ stand for the spin-orbit wave functions for the initial and finial states, respectively. For example, the matrix elements $\langle f|\bm{\epsilon}_2\cdot\bm{\epsilon_4^{\dag}}|i\rangle$ for the $J^P=1/2^-$ and $3/2^-$ $\Xi_c\bar{D}^*$ systems are $\text{diag}(1,1,1)$ and $\text{diag}(1,1,1,1)$ respectively when input the spin-orbit wave functions in Eq. (\ref{spinwave}). As we seen, the OBE effective potentials are exactly the same for the $\Xi_c\bar{D}^*$ systems with $J^P=1/2^-$ and $3/2^-$.

\begin{figure}[htbp]
  \centering
  \includegraphics[width=0.46\textwidth]{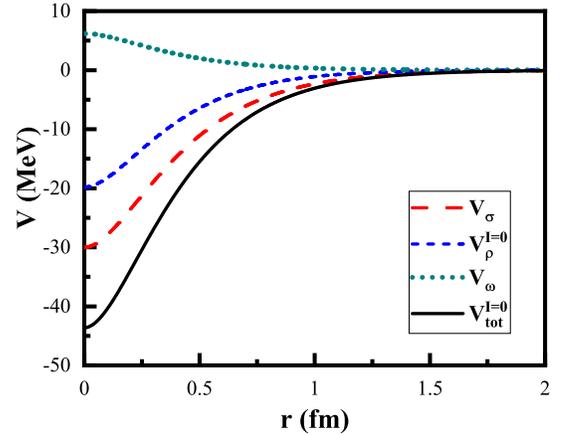}\\
\caption{The dependence of the $S-$wave OBE effective potentials for the isoscalar $\Xi_c\bar{D}^*$ system on $r$ with $\Lambda=1.00$ GeV.}\label{Poten1}
\end{figure}

There exist $\sigma$, $\rho$, and $\omega$ exchanges interactions for the single $\Xi_c\bar{D}^*$ system in Eq. (\ref{pot1}). In Fig. \ref{Poten1}, we present the corresponding $S-$wave effective potentials for the single $\Xi_c\bar{D}^*$ system with $I(J^P)=0(1/2^-,3/2^-)$. Here, the $\sigma$ and $\omega$ exchanges provide an attractive and a repulsive interactions, respectively. The $\rho$ exchange interaction is very different for the isoscalar and isovector $\Xi_c\bar{D}^*$ systems, i.e., for the isovector case, the $\rho$ exchange provides a repulsive interaction, whereas a three times stronger attractive force in the isoscalar case.

\begin{figure}[htbp]
  \centering
  \includegraphics[width=0.46\textwidth]{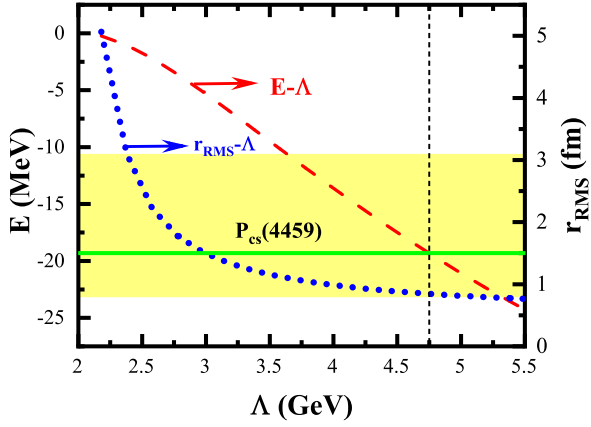}\\
\caption{The $\Lambda$ dependence of the bound solutions (the binding energy $E$ and the root-mean-square radius $r_{RMS}$) for the single $\Xi_c\bar{D}^*$ states. Here, the dash area corresponds to the $P_{cs}(4459)$ mass with experimental uncertainties. The horizontal solid line and the vertical dotted line stand for the central mass of the $P_{cs}(4459)$ and the corresponding cutoff value.}\label{Er1}
\end{figure}

After solving the Shr\"{o}dinger equation, we find that bound solutions for the isoscalar $\Xi_c\bar{D}^*$ systems with $J^P=1/2^-(3/2^-)$ appear in the range of $\Lambda>2.00$ GeV. With the increasing of the cutoff $\Lambda$, it binds more strongly. As shown in Fig. \ref{Er1}, when the cutoff $\Lambda$ is taken in the range of $3.6<\Lambda<5.25$ GeV, we can reproduce the mass of the $P_{cs}(4459)$ with the experimental uncertainties. Obviously, the cutoff $\Lambda$ is far away from the typical value 1.00 GeV in a loosely bound hadronic molecular state. Thus, our results indicate that the newly $P_{cs}(4459)$ cannot be a pure $\Xi_c\bar{D}^*$ molecule, although the total OBE effective potential is attractive.

Compared to the isoscalar $\Xi_c\bar{D}^*$ system, the OBE effective potential for the isovector system is much weaker attractive. We don't find the bound solutions in the range of $\Lambda<5.00$ GeV.

\section{A coupled $\Xi_c\bar{D}^*/\Xi_c^*\bar{D}/\Xi_c^{\prime}\bar{D}^*/\Xi_c^*\bar{D}^*$ analysis}\label{sec4}

In this section, we further introduce the coupled channel effect to check whether the $P_{cs}(4459)$ can be explained as a strange hidden-charm molecular pentaquark. Here, we study the $\Xi_c\bar{D}^*/\Xi_c^*\bar{D}/\Xi_c^{\prime}\bar{D}^*/\Xi_c^*\bar{D}^*$ interactions with $I(J^P)=0,1(1/2^-,3/2^-)$. The total effective potentials are
\begin{eqnarray}
{V}  &=& \left(\begin{array}{cccc}
\mathcal{V}^{11}   &\mathcal{V}^{12}    &\mathcal{V}^{13}    &\mathcal{V}^{14}\\
\mathcal{V}^{21}   &\mathcal{V}^{22}    &\mathcal{V}^{23}    &\mathcal{V}^{24}\\
\mathcal{V}^{31}   &\mathcal{V}^{32}    &\mathcal{V}^{33}    &\mathcal{V}^{34}\\
\mathcal{V}^{41}   &\mathcal{V}^{42}    &\mathcal{V}^{43}    &\mathcal{V}^{44}\end{array}\right)\nonumber\\
&=& {\left(\begin{array}{cccc}
\mathcal{V}^{\Xi_c\bar{D}^*\to\Xi_c\bar{D}^*}
              &\mathcal{V}^{\Xi_c^*\bar{D}\to\Xi_c\bar{D}^*}
              &\mathcal{V}^{\Xi_c^{\prime}\bar{D}^*\to\Xi_c\bar{D}^*}
              &\mathcal{V}^{\Xi_c^*\bar{D}^*\to\Xi_c\bar{D}^*}\\
\mathcal{V}^{\Xi_c\bar{D}^*\to\Xi_c^*\bar{D}}
              &\mathcal{V}^{\Xi_c^*\bar{D}\to\Xi_c^*\bar{D}}
              &\mathcal{V}^{\Xi_c^{\prime}\bar{D}^*\to\Xi_c^*\bar{D}}
              &\mathcal{V}^{\Xi_c^*\bar{D}^*\to\Xi_c^*\bar{D}}\\
\mathcal{V}^{\Xi_c\bar{D}^*\to\Omega_c^*\eta}
              &\mathcal{V}^{\Xi_c^*\bar{D}\to\Xi_c^{\prime}\bar{D}^*}
              &\mathcal{V}^{\Xi_c^{\prime}\bar{D}^*\to\Xi_c^{\prime}\bar{D}^*}
              &\mathcal{V}^{\Xi_c^*\bar{D}^*\to\Xi_c^{\prime}\bar{D}^*}\\
\mathcal{V}^{\Xi_c\bar{D}^*\to\Xi_c^*\bar{D}^*}
              &\mathcal{V}^{\Xi_c^*\bar{D}\to\Xi_c^*\bar{D}^*}
              &\mathcal{V}^{\Xi_c^{\prime}\bar{D}^*\to\Xi_c^*\bar{D}^*}
              &\mathcal{V}^{\Xi_c^*\bar{D}^*\to\Xi_c^*\bar{D}^*}\end{array}\right)}.\nonumber\\\label{hh}
\end{eqnarray}
The subpotentials are expressed as
\begin{eqnarray}
\mathcal{V}^{11} &=& 2\mathbb{A}\mathcal{Y}^{11}_{\Lambda,m_{\sigma}}
    -\frac{\mathbb{B}}{4}\left(\mathcal{G}(I)\mathcal{Y}^{11}_{\Lambda,m_{\rho}}+\mathcal{Y}^{11}_{\Lambda,m_{\omega}}\right),\\
\mathcal{V}^{12} &=&
   -\frac{\mathbb{C}}{6\sqrt{2}}\left(\mathcal{G}(I)\mathcal{Z}^{12}_{\Lambda_1,m_{\pi1}}
   +\mathcal{Z}^{12}_{\Lambda_1,m_{\eta1}}\right)\nonumber\\
   &&-\frac{\mathbb{D}}{6\sqrt{2}}\left(\mathcal{G}(I)\mathcal{Z}^{\prime12}_{\Lambda_1,m_{\rho1}}
   +\mathcal{Z}^{\prime12}_{\Lambda_1,m_{\omega1}}\right),\\
\mathcal{V}^{13} &=&
   \frac{\mathbb{C}}{6\sqrt{6}}\left(\mathcal{G}(I)\mathcal{Z}^{13}_{\Lambda_2,m_{\pi2}}
   +\mathcal{Z}^{13}_{\Lambda_2,m_{\eta2}}\right)\nonumber\\
   &&
   -\frac{\mathbb{D}}{3\sqrt{6}}\left(\mathcal{G}(I)\mathcal{Z}^{\prime13}_{\Lambda_2,m_{\rho2}}
   +\mathcal{Z}^{\prime13}_{\Lambda_2,m_{\omega2}}\right),\\
\mathcal{V}^{14} &=&
   -\frac{\mathbb{C}}{6\sqrt{2}}\left(\mathcal{G}(I)\mathcal{Z}^{14}_{\Lambda_3,m_{\pi3}}
   +\mathcal{Z}^{14}_{\Lambda_3,m_{\eta3}}\right)\nonumber\\
   &&
   +\frac{\mathbb{D}}{3\sqrt{2}}\left(\mathcal{G}(I)\mathcal{Z}^{\prime14}_{\Lambda_3,m_{\rho3}}
   +\mathcal{Z}^{\prime14}_{\Lambda_3,m_{\omega3}}\right),\\
\mathcal{V}^{22} &=&
    -\mathbb{A}^{\prime}\mathcal{Y}^{22}_{\Lambda,m_{\sigma}}
    +\frac{\mathbb{B}^{\prime}}{8}\left(\mathcal{G}(I)\mathcal{Y}^{22}_{\Lambda,m_{\rho}}
    +\mathcal{Y}^{22}_{\Lambda,m_{\omega}}\right),\\
\mathcal{V}^{23}&=&
    \frac{\mathbb{C}^{\prime}}{8\sqrt{3}}\left(\mathcal{G}(I)\mathcal{Z}^{23}_{\Lambda_{4},m_{\pi{4}}}
    -\frac{1}{3}\mathcal{Z}^{23}_{\Lambda_{4},m_{\eta{4}}}\right)\nonumber\\
   &&
    -\frac{\mathbb{D}^{\prime}}{12\sqrt{3}}\left(\mathcal{G}(I)\mathcal{Z}^{\prime23}_{\Lambda_{4},m_{\rho{4}}}
    +\mathcal{Z}^{\prime23}_{\Lambda_{4},m_{\omega{4}}}\right),\\
\mathcal{V}^{24} &=&
    \frac{\mathbb{C}^{\prime}}{8}\left(\mathcal{G}(I)\mathcal{Z}^{24}_{\Lambda_{5},m_{\pi{5}}}
    -\frac{1}{3}\mathcal{Z}^{24}_{\Lambda_{5},m_{\eta{5}}}\right)\nonumber\\
   &&
    -\frac{\mathbb{D}^{\prime}}{12}\left(\mathcal{G}(I)\mathcal{Z}^{\prime24}_{\Lambda_{5},m_{\rho{5}}}
    +\mathcal{Z}^{\prime 24}_{\Lambda_{5},m_{\omega{5}}}\right),\\
\mathcal{V}^{33} &=&
     -\mathbb{A}^{\prime}\mathcal{Y}^{33}_{\Lambda,m_{\sigma}}
     +\frac{\mathbb{C}^{\prime}}{12}\left(\mathcal{G}(I)\mathcal{Z}^{33}_{\Lambda,m_{\pi}}
     -\frac{1}{3}\mathcal{Z}^{33}_{\Lambda,m_{\eta}}\right)\nonumber\\
   &&
    -\frac{\mathbb{B}^{\prime}}{8}\left(\mathcal{G}(I)\mathcal{Y}^{33}_{\Lambda,m_{\rho}}
    +\mathcal{Y}^{33}_{\Lambda,m_{\omega}}\right)\nonumber\\
   &&
    -\frac{\mathbb{D}^{\prime}}{18}\left(\mathcal{G}(I)\mathcal{Z}^{\prime33}_{\Lambda,m_{\rho}}
    +\mathcal{Z}^{\prime33}_{\Lambda,m_{\omega}}\right),\\
\mathcal{V}^{34} &=&
    \frac{\mathbb{A}^{\prime}}{\sqrt{3}}\mathcal{Y}^{34}_{\Lambda_6,m_{\sigma6}}
     +\frac{\mathbb{C}^{\prime}}{8\sqrt{3}}\left(\mathcal{G}(I)\mathcal{Z}^{34}_{\Lambda_6,m_{\pi6}}
     -\frac{1}{3}\mathcal{Z}^{34}_{\Lambda_6,m_{\eta6}}\right)\nonumber\\
   &&
    -\frac{\mathbb{B}^{\prime}}{8\sqrt{3}}\left(\mathcal{G}(I)\mathcal{Y}^{34}_{\Lambda_6,m_{\rho6}}
       +\mathcal{Y}^{34}_{\Lambda_6,m_{\omega6}}\right)\nonumber\\
   &&
    -\frac{\mathbb{D}^{\prime}}{12\sqrt{3}}\left(\mathcal{G}(I)\mathcal{Z}^{\prime34}_{\Lambda_6,m_{\rho6}}
    +\mathcal{Z}^{\prime34}_{\Lambda_6,m_{\omega6}}\right),\\
\mathcal{V}^{44} &=&
     -\mathbb{A}^{\prime}\mathcal{Y}^{44}_{\Lambda,m_{\sigma}}
     -\frac{\mathbb{C}^{\prime}}{8}\left(\mathcal{G}(I)\mathcal{Z}^{44}_{\Lambda,m_{\pi}}
     -\frac{1}{3}\mathcal{Z}^{44}_{\Lambda,m_{\eta}}\right)\nonumber\\
   &&
    -\frac{\mathbb{B}^{\prime}}{8}\left(\mathcal{G}(I)\mathcal{Y}^{44}_{\Lambda,m_{\rho}}
    +\mathcal{Y}^{44}_{\Lambda,m_{\omega}}\right)\nonumber\\
   &&
    -\frac{\mathbb{D}^{\prime}}{12}\left(\mathcal{G}(I)\mathcal{Z}^{\prime44}_{\Lambda,m_{\rho}}
    +\mathcal{Z}^{\prime44}_{\Lambda,m_{\omega}}\right).
\end{eqnarray}
In the above expressions, we define several useful functions, i.e.,
\begin{eqnarray}
\mathcal{Y}^{ij}_{\Lambda, m_a}&=&\mathcal{D}_{ij}Y(\Lambda,m_\sigma,r),\\
\mathcal{Z}^{ij}_{\Lambda, m_a}&=&\left(\mathcal{E}_{ij}\nabla^2+\mathcal{F}_{ij}r\frac{\partial}{\partial r}\frac{1}{r}\frac{\partial}{\partial r}\right)Y(\Lambda,m_a,r),\\
\mathcal{Z}^{\prime ij}_{\Lambda,
m_a}&=&\left(2\mathcal{E}_{ij}\nabla^2-\mathcal{F}_{ij}r\frac{\partial}{\partial
r}\frac{1}{r}\frac{\partial}{\partial r}\right)Y(\Lambda,m_a,r).
\end{eqnarray}
$\mathcal{D}_{ij}$, $\mathcal{E}_{ij}$, and $\mathcal{F}_{ij}$ stand for the spin-spin interaction and tensor force operators, the concrete expressions are summarized in Appendix. The variables in these functions are defined as $\Lambda_i^2 =\Lambda^2-q_i^2$, $m_{{i}}^2=m^2-q_i^2$, with $i=0$, 1, ..., 6. The coupling parameters and the values for $q_i$ are summarized in Table \ref{parameter}.

\renewcommand\tabcolsep{0.3cm}
\renewcommand{\arraystretch}{1.7}
\begin{table}[!htbp]
\caption{Coupling parameters and the values for $q_i$ adopted in our calculation. The unite for $q_i$ is GeV.\label{parameter}}
\begin{tabular}{cccc}
\toprule[1pt]
    $\mathbb{A}=l_Bg_s$     &$\mathbb{B}=\beta\beta_Bg_v^2$     &$\mathbb{C}=gg_4/f_{\pi}^2$    &$\mathbb{D}=\lambda\lambda_Ig_v^2$\\
    $\mathbb{A}^{\prime}=l_sg_s$     &$\mathbb{B}^{\prime}=\beta\beta_sg_v^2$     &$\mathbb{C}^{\prime}=gg_1/f_{\pi}^2$    &$\mathbb{D}^{\prime}=\lambda\lambda_sg_v^2$\\
    $q_1=0.16$      &$q_2=0.06$     &$q_3=0.10$\\
    $q_4=0.10$      &$q_5=0.06$     &$q_6=0.04$\\\bottomrule[1pt]
\end{tabular}
\end{table}

When producing the numerical calculation, we neglect the $D-$wave components for all discussed systems. After solving the coupled channel Shr$\ddot{\text{o}}$dinger equation, we can only obtain bound states with their masses below the lowest threshold of the discussed channels. Here, we need to mention that one can further search for the possible resonances with higher masses by solving the scattering problems, which we leave to the next task. As we will see in the following results, we also take the cutoff value of cutoff from 1.00 GeV to 5.00 GeV (see Table \ref{num1} for more details).
\renewcommand\tabcolsep{0.12cm}
\renewcommand{\arraystretch}{1.8}
\begin{table}[!htbp]
  \centering
  \caption{The obtained bound state solutions (the binding energy $E$, the root-mean-square $r_{RMS}$, and the probabilities of different channels for the investigated systems) of the coupled $\Xi_c\bar{D}^*/\Xi_c^*\bar{D}/\Xi_c^{\prime}\bar{D}^*/\Xi_c^*\bar{D}^*$ systems with $I(J^P)=0(1/2^-,3/2^-)$. Here, the cutoff $\Lambda$, the root-mean-square $r_{RMS}$, and the binding energy $E$ are in the units of GeV, fm, and MeV, respectively. The dominant channels for a bound state are remarked by bold typeface.}\label{num1}
  \begin{tabular}{c|ccc|cccc}
    \toprule[1pt]
  $I(J^P)$     &$\Lambda$     &$E$     &$r_{RMS}$
           &$\Xi_c\bar{D}^*$     &$\Xi_c^*\bar{D}$    &$\Xi_c^{\prime}\bar{D}^*$    &$\Xi_c^{*}\bar{D}^*$\\\hline
  $0(1/2^-)$   &1.17    &$-1.63$     &1.39      &30.66     &$\ldots$       &$\bm{64.13}$      &5.21\\
               &1.18    &$-7.52$     &0.62      &15.82     &$\ldots$       &$\bm{77.10}$       &7.08\\
               &1.19    &$-14.29$    &0.50      &11.12     &$\ldots$       &$\bm{80.82}$       &8.06\\
               &1.20    &$-21.62$    &0.45      &8.60      &$\ldots$       &$\bm{82.62}$       &8.78\\\hline

  $0(3/2^-)$   &0.99    &$-1.46$     &2.18      &$\bm{69.44}$     &19.46      &2.81      &8.28\\
               &1.01    &$-5.73$     &1.09      &$\bm{53.70}$     &28.41      &5.52      &13.37\\
               &1.03    &$-11.77$    &0.79      &$\bm{44.88}$     &$\bm{32.50}$   &5.69    &16.93\\
               &1.05    &$-19.28$    &0.65      &$\bm{38.95}$     &$\bm{34.58}$   &6.61    &18.86\\
  \bottomrule[1pt]
  \end{tabular}
\end{table}

For the $I(J^P)=0(3/2^-)$ case, there exist bound state solutions with the cutoff around 1.00 GeV. And the $\Xi_c\bar{D}^*$ channel has a dominant contribution. It is obvious that the value of the cutoff for the coupled channel case is smaller than that for the single channel case. Thus, we can conclude that the coupled channel effect is helpful to form this bound state. We also notice that its root-mean-square (RMS) radius is around 1 fm when the binding energy varies from $0$ to $-10$ MeV, the size is reasonable for such a coupled $\Xi_c\bar{D}^*/\Xi_c^*\bar{D}/\Xi_c^{\prime}\bar{D}^*/\Xi_c^*\bar{D}^*$ molecular system. As it binds deeper and deeper, the $\Xi_c^*\bar{D}$ component becomes more and more important. Especially, when we reproduce the mass of $P_{cs}(4459)$, the probabilities for both $\Xi_c\bar{D}^*$ and $\Xi_c^*\bar{D}$ channels are over 30 percent.

When the cutoff is taking from 1.17 GeV to 1.20 GeV, we can obtain a loosely bound state for the $I(J^P)=0(1/2^-)$ case. Here, the $\Xi_c^{\prime}\bar{D}^*$ channel is its dominant channel. The "real" binding energy is over 100 MeV in the range of $1.17<\Lambda<1.20$ GeV. Where the "real" binding energy $E_{rel}$ is $E_{rel}=E+M_{lowest}-M_{dominant}$, $M_{lowest}$ and $M_{dominant}$ are the mass thresholds of the channel with lowest mass and the dominant channel, respectively. Although the cutoff is reasonable, the corresponding RMS radius is only around or less than 0.5 fm when reproducing the mass of $P_{cs}(4459)$. It is obvious that the molecular state picture with the $I(J^P)=0(1/2^-)$ assignment to $P_{cs}(4459)$ is not favored.

To summarize, the coupled $\Xi_c\bar{D}^*/\Xi_c^*\bar{D}/\Xi_c^{\prime}\bar{D}^*/\Xi_c^*\bar{D}^*$ state with $I(J^P)=0(3/2^-)$ can be a possible strange hidden-charm molecular candidate, the $\Xi_c\bar{D}^*$ and $\Xi_c^*\bar{D}$ channels are dominant, followed by the $\Xi_c^*\bar{D}^*$ channel. It may relate to the newly $P_{cs}(4459)$ recently reported by the LHCb Collaboration. Here, we also obtain bound state solutions for the coupled $\Xi_c\bar{D}^*/\Xi_c^*\bar{D}/\Xi_c^{\prime}\bar{D}^*/\Xi_c^*\bar{D}^*$ system with $I(J^P)=0(1/2^-)$, its "real" binding energy reaches around 100 MeV as the dominant channel is the $\Xi_c^{\prime}\bar{D}^*$ channel, thus, the relevant OBE interaction from the $\Xi_c^{\prime}\bar{D}^*$ system is strongly attractive. Although it isn't a reasonable loose hadronic molecular candidate, in our future study, we can search for possible loosely bound strange hidden-charm molecules near the threshold of the $\Xi_c^{\prime}\bar{D}^*$ system.

\renewcommand\tabcolsep{0.12cm}
\renewcommand{\arraystretch}{1.8}
\begin{table}[!htbp]
  \centering
  \caption{The obtained bound state solutions (the binding energy $E$, the root-mean-square $r_{RMS}$, and the probabilities of different channels for the investigated system) of the coupled $\Xi_c\bar{D}^*/\Xi_c^*\bar{D}/\Xi_c^{\prime}\bar{D}^*/\Xi_c^*\bar{D}^*$ systems with $I(J^P)=1(1/2^-,3/2^-)$. Here, the cutoff $\Lambda$, the root-mean-square $r_{RMS}$, and the binding energy $E$ are in the units of GeV and fm, and MeV, respectively. The dominant channels for a bound state are remarked by bold typeface.}\label{num2}
  \begin{tabular}{c|ccc|cccc}
    \toprule[1pt]
  $I(J^P)$     &$\Lambda$     &$E$     &$r_{RMS}$
           &$\Xi_c\bar{D}^*$     &$\Xi_c^*\bar{D}$    &$\Xi_c^{\prime}\bar{D}^*$    &$\Xi_c^{*}\bar{D}^*$\\\hline
  $1(1/2^-)$   &2.07    &$-2.05$     &0.36      &5.80      &$\ldots$       &5.33     &$\bm{88.87}$ \\
               &2.08    &$-16.84$    &0.24      &5.03      &$\ldots$       &5.30     &$\bm{89.67}$ \\
               &2.09    &$-32.20$    &0.23      &4.95      &$\ldots$       &5.23     &$\bm{89.81}$ \\
               &2.10    &$-48.12$    &0.22      &4.95      &$\ldots$       &5.17     &$\bm{89.88}$ \\\hline
  $1(3/2^-)$   &1.43    &$-1.62$     &0.94      &12.96      &$\bm{80.10}$     &0.61      &6.34\\
               &1.44    &$-5.68$     &0.67      &9.92       &$\bm{82.29}$     &0.64      &7.14\\
               &1.45    &$-10.12$    &0.60      &9.12       &$\bm{82.41}$     &0.66      &7.81\\
               &1.46    &$-14.91$    &0.57      &8.78       &$\bm{82.08}$     &0.67      &8.47\\
  \bottomrule[1pt]
  \end{tabular}
\end{table}

As a byproduct, we further discuss the existence of the isovector strange hidden-charm molecular pentaquarks from the $S-$wave $\Xi_c\bar{D}^*/\Xi_c^*\bar{D}/\Xi_c^{\prime}\bar{D}^*/\Xi_c^*\bar{D}^*$ interactions. According to the numerical results in Table \ref{num2}, we find the $I(J^P)=1(1/2^-)$ $\Xi_c\bar{D}^*/\Xi_c^*\bar{D}/\Xi_c^{\prime}\bar{D}^*/\Xi_c^*\bar{D}^*$ state is a tight bound state, the $\Xi_c^*\bar{D}^*$ channel is the dominant channel. However, we need to specify that the cutoff $\Lambda$ is litter bit larger than 1 GeV. For the $I(J^P)=1(3/2^-)$ case, when the cutoff reaches up to 1.43 GeV, we can obtain a bound state, where the $\Xi_c^*\bar{D}$ channel is dominant. If strictly taking $\Lambda$ around $1$ GeV and the RMS radius around $1$ fm, we may predict there is a possible strange hidden-charm $\Xi_c^*\bar{D}$ molecular candidate.

\section{Summary}\label{sec5}

Inspired by the evidence of the first strange hidden-charm pentaquark $P_{cs}(4459)$, we systematically discuss whether the $P_{cs}(4459)$ can be assigned as the $\Xi_c\bar{D}^*$ molecular state. In this work, we adopt the OBE effective potentials and present our results in two cases.

In case one, we study the single $\Xi_c\bar{D}^*$ system and consider the $S-D$ wave mixing effect. The long range interaction from the one-$\pi$-exchange is suppressed by the spin-parity forbidden. There remain the intermediate range force from the one-$\sigma$-exchange and the short range force from the one-$\rho/\omega-$exchange. The OBE effective potentials for the $I(J^P)=0(1/2^-,3/2^-)$ are the exactly same. Our results indicates the OBE effective potentials are not strong enough to form a single $\Xi_c\bar{D}^*$ system.

In case two, we further study the coupled $\Xi_c\bar{D}^*/\Xi_c^*\bar{D}/\Xi_c^{\prime}\bar{D}^*/\Xi_c^*\bar{D}^*$ systems with $I(J^P)=0(1/2^-,3/2^-)$ after introducing the coupled channel effect. Finally, we obtain a good strange hidden-charm molecular pentaquark with $I(J^P)=0(3/2^-)$, it is mainly composed by the $\Xi_c\bar{D}^*$ and $\Xi_c^*\bar{D}$ channels. The coupled channel effect plays an important role here. When the cutoff is taken as $\Lambda=1.05$ GeV, we can reproduce the central mass of the $P_{cs}(4459)$, and the probabilities for the $\Xi_c\bar{D}^*$, $\Xi_c^*\bar{D}$, $\Xi_c^{\prime}\bar{D}^*$, and $\Xi_c^*\bar{D}^*$ channels are 38.95\%, 34.58\%, 6.61\%, and 18.86\%, respectively.

In addition, we also find the OBE effective potentials for the $\Xi_c^{\prime}\bar{D}^*$ system with $0(1/2^-)$ and the $\Xi_c^{*}\bar{D}$ system with $1(3/2^-)$ are strongly attractive in the reasonable cutoff input, where one may search for more possible strange hidden-charm molecules or resonances.

\appendix*

\section{Operators}\label{app}
The concrete operators in the OBE effective potentials are
\begin{eqnarray*}
\mathcal{D}_{11} &=& \chi_3^{\dag}\chi_1\bm{\epsilon}_2\cdot\bm{\epsilon}_4^{\dag},\\
\mathcal{E}_{12} &=& \sum_{m,n}C_{1/2,m;1,n}^{3/2,m+n}\chi_{3,m}^{\dag}
    \bm{\epsilon}_{3,n}^{\dag}\cdot\bm{\epsilon}_{2}\chi_1,\\
\mathcal{F}_{12} &=& \sum_{m,n}C_{1/2,m;1,n}^{3/2,m+n}\chi_{3,m}^{\dag}
    S\left(\hat{r},\bm{\epsilon}_{3,n}^{\dag},\bm{\epsilon}_{2}\right)\chi_1,\\
\mathcal{E}_{13} &=& \chi_3^{\dag}\bm{\sigma}\cdot\left(i\bm{\epsilon}_2\times
     \bm{\epsilon}_4^{\dag}\right)\chi_1,\\
\mathcal{F}_{13} &=& \chi_3^{\dag}S\left(\hat{r},\bm{\sigma},
     i\bm{\epsilon}_2\times\bm{\epsilon}_4^{\dag}\right)\chi_1,\\
\mathcal{E}_{14} &=& \sum_{m,n}C_{1/2,m;1,n}^{3/2,m+n}\chi_{3,m}^{\dag}
       \left(i\bm{\epsilon}_2\times\bm{\epsilon}_4^{\dag}\right)\cdot\bm{\epsilon}_{3,n}^{\dag}\chi_1,\\
\mathcal{F}_{14} &=& \sum_{m,n}C_{1/2,m;1,n}^{3/2,m+n}\chi_{3,m}^{\dag}
       S\left(\hat{r},\bm{\epsilon}_{3,n}^{\dag},i\bm{\epsilon}_2\times\bm{\epsilon}_4^{\dag}\right)\chi_1,\\
\mathcal{D}_{22} &=& \sum_{a,b}^{m,n}C_{1/2,a;1,b}^{3/2,a+b}
   C_{1/2,m;1,n}^{3/2,m+n}\chi_3^{a\dag}\chi_1^m\bm{\epsilon}_1^n\cdot\bm{\epsilon}_3^{b\dag},\\
\mathcal{E}_{32} &=& \sum_{m,n}C_{1/2,m;1,n}^{3/2,m+n}\chi_3^{\dag}
     \bm{\epsilon}_4^{\dag}\cdot\left(i\bm{\sigma}\times\bm{\epsilon}_1^{n}\right)\chi_{1,m},\\
\mathcal{F}_{32} &=& \sum_{m,n}C_{1/2,m;1,n}^{3/2,m+n}\chi_3^{\dag}
     S\left(\hat{r},\bm{\epsilon}_4^{\dag},i\bm{\sigma}\times\bm{\epsilon}_1^{n}\right)\chi_{1,m},\\
\mathcal{E}_{42} &=& \sum_{a,b}^{m,n}C_{1/2,a;1,b}^{3/2,a+b}
     C_{1/2,m;1,n}^{3/2,m+n}\chi_3^{a\dag}\chi_1^{m}
     \bm{\epsilon}_4^{\dag}\cdot\left(i\bm{\epsilon}_1^{n}\times
                 \bm{\epsilon}_3^{b\dag}\right),\\
\mathcal{F}_{42} &=& \sum_{a,b}^{m,n}C_{1/2,a;1,b}^{3/2,a+b}
     C_{1/2,m;1,n}^{3/2,m+n}\chi_3^{a\dag}\chi_1^{m}
     S\left(\hat{r},\bm{\epsilon}_4^{\dag},i\bm{\epsilon}_1^{n}\times\bm{\epsilon}_3^{b\dag}\right),\\
\mathcal{D}_{33} &=& \chi_3^{\dag}\chi_1\bm{\epsilon}_2\cdot\bm{\epsilon}_4^{\dag},\\
\mathcal{E}_{33} &=& \chi_3^{\dag}\bm{\sigma}\cdot\left(i\bm{\epsilon}_2\times
     \bm{\epsilon}_4^{\dag}\right)\chi_1,\\
\mathcal{F}_{33} &=& \chi_3^{\dag}S\left(\hat{r},\bm{\sigma},
     i\bm{\epsilon}_2\times\bm{\epsilon}_4^{\dag}\right)\chi_1,\\
\mathcal{D}_{43} &=& \sum_{m,n}C_{\frac{1}{2},m;1,n}^{\frac{3}{2},m+n}\chi_3^{m\dag}
    \left(\bm{\sigma}\cdot\bm{\epsilon}_3^{n\dag}\right)
    \left(\bm{\epsilon}_2\cdot\bm{\epsilon}_4^{\dag}\right)\chi_1,\\
\mathcal{E}_{43} &=& \sum_{m,n}C_{\frac{1}{2},m;1,n}^{\frac{3}{2},m+n}\chi_3^{m\dag}
    \left(\bm{\sigma}\times\bm{\epsilon}_3^{n\dag}\right)\cdot
    \left(\bm{\epsilon}_2\times\bm{\epsilon}_4^{\dag}\right)\chi_1,\\
\mathcal{F}_{43} &=& \sum_{m,n}C_{\frac{1}{2},m;1,n}^{\frac{3}{2},m+n}\chi_3^{m\dag}
    S\left(\hat{r},\bm{\sigma}\times\bm{\epsilon}_3^{n\dag},
    \bm{\epsilon}_2\times\bm{\epsilon}_4^{\dag}\right)\chi_1,\\
\mathcal{D}_{44} &=& \sum_{a,b}^{m,n}C_{\frac{1}{2},a;1,b}^{\frac{3}{2},a+b}
    C_{\frac{1}{2},m;1,n}^{\frac{3}{2},m+n}\chi_3^{a\dag}
    \left(\bm{\epsilon}_1^n\cdot\bm{\epsilon}_3^{b\dag}\right)
    \left(\bm{\epsilon}_2\cdot\bm{\epsilon}_4^{\dag}\right)\chi_1^m,\\
\mathcal{E}_{44} &=& \sum_{a,b}^{m,n}C_{\frac{1}{2},a;1,b}^{\frac{3}{2},a+b}
    C_{\frac{1}{2},m;1,n}^{\frac{3}{2},m+n}\chi_3^{a\dag}
    \left(\bm{\epsilon}_1^n\times\bm{\epsilon}_3^{b\dag}\right)
    \cdot\left(\bm{\epsilon}_2\times\bm{\epsilon}_4^{\dag}\right)\chi_1^m,\\
\mathcal{F}_{44} &=& \sum_{a,b}^{m,n}C_{\frac{1}{2},a;1,b}^{\frac{3}{2},a+b}
    C_{\frac{1}{2},m;1,n}^{\frac{3}{2},m+n}\chi_3^{a\dag}
    S\left(\hat{r},\bm{\epsilon}_1^n\times\bm{\epsilon}_3^{b\dag},
    \bm{\epsilon}_2\times\bm{\epsilon}_4^{\dag}\right)\chi_1^m.
\end{eqnarray*}

\section*{ACKNOWLEDGMENTS}

Rui Chen is very grateful to Xiang Liu and Shi-Lin Zhu for helpful discussions and constructive suggestions. This project is supported by the National Postdoctoral Program for Innovative Talent.

\end{document}